\documentclass[aps,prc,twocolumn,nofootinbib]{revtex4-1}
\usepackage[utf8]{inputenc}
\usepackage{csquotes}
\usepackage{todonotes}
\usepackage{amsmath,amssymb}
\usepackage{physics}
\usepackage{units}
\usepackage{bbold}
\usepackage{textcomp}
\usepackage{hyperref}

\newcommand{\chieft}{$\chi$EFT}
\newcommand{\NN}{$NN$}
\newcommand{\NNN}{$NNN$}

\newcommand{\np}{$np$}
\newcommand{\pp}{$pp$}
\newcommand{\piN}{$\pi N$}
\newcommand{\Ctnn}{
	\ifmmode \widetilde{C}_{1S0}^{nn} \else$\widetilde{C}_{1S0}^{nn}$\fi
}
\newcommand{\Ctnp}{
	\ifmmode \widetilde{C}_{1S0}^{np} \else$\widetilde{C}_{1S0}^{np}$\fi
}
\newcommand{\Ctpp}{
	\ifmmode \widetilde{C}_{1S0}^{pp} \else$\widetilde{C}_{1S0}^{pp}$\fi
}
\newcommand{\prob}{\textnormal{pr}}
\newcommand{\lec}{\alpha}
\newcommand{\lecs}{\vec{\lec}}
\newcommand{\pinlecs}{\vec{\lec}_{\pi N}}
\newcommand{\nnlecs}{\vec{\lec}_{NN}}
\newcommand{\pdf}{PDF}
\newcommand{\ppd}{PPD}
\newcommand{\cov}{\Sigma}
\newcommand{\covmtx}{\boldsymbol{\cov}}
\newcommand{\data}{\boldsymbol{\mathcal{D}}}
\newcommand{\tlab}{T_\text{lab}}
\newcommand{\Ct}{\widetilde{C}}

\newcommand{\yexp}{y_\text{exp}}
\newcommand{\yth}{y_\text{th}}
\newcommand{\yref}{y_\text{ref}}
\newcommand{\ystar}{y^\star}
\newcommand{\dyexp}{\delta y_\text{exp}}
\newcommand{\dyth}{\delta y_\text{th}}
\newcommand{\idpt}{\vec{x}}
\newcommand{\idptstar}{\idpt^\star}
\newcommand{\peth}{\boldsymbol{x}}
\newcommand{\cbar}{\bar{c}}
\newcommand{\cbarsq}{\cbar^2}
\newcommand{\vecc}{\boldsymbol{c}}
\newcommand{\ctrain}{\vecc^{\star}}
\newcommand{\cval}{\widetilde{\vecc}}

\newcommand{\xval}{\widetilde{\peth}}

\newcommand{\ls}{\boldsymbol{\ell}}
\newcommand{\lsth}{\ell_{\theta}}
\newcommand{\lsE}{\ell_{\tlab}}
\newcommand{\nobs}{N_y}
\newcommand{\nval}{N_\text{val}}
\newcommand{\ndatay}{N_{d,y}}
\newcommand{\nE}{N_{\tlab}}
\newcommand{\nEy}{N_{\tlab,y}}
\newcommand{\GP}{\mathcal{GP}}
\newcommand{\gpcovval}{\widetilde{\boldsymbol{\Sigma}}_{\GP}}

\newcommand{\ckernel}{\kappa}
\newcommand{\ykernel}{K}
\newcommand{\given}{\, | \,}
\newcommand{\normeig}{\gamma}

\newcommand{\mdsquared}{D^2_\textnormal{MD}}
\newcommand{\Ctunit}{
  \ifmmode \times 10^4\,\text{GeV}^{-2} \else $\times 10^4\,$GeV$^{-2}$ \fi
}
\newcommand{\Cunit}{
  \ifmmode \times 10^4\,\text{GeV}^{-4} \else $\times 10^4\,$GeV$^{-4}$ \fi
}
\newcommand{\Dunit}{
  \ifmmode \times 10^4\,\text{GeV}^{-6} \else $\times 10^4\,$GeV$^{-6}$ \fi
}
\newcommand{\cunit}{
  \ifmmode \text{ GeV}^{-1} \else GeV$^{-1}$ \fi
}
\newcommand{\Tlab}{
  \ifmmode T_{\textnormal{lab}} \else $T_{\textnormal{lab}}$ \fi
}
\newcommand{\lo}{
  \ifmmode \textnormal{LO} \else\unskip LO\fi
}
\newcommand{\nlo}{
  \ifmmode \textnormal{NLO} \else\unskip NLO\fi
}
\newcommand{\nnlo}{
  \ifmmode \textnormal{NNLO} \else\unskip NNLO\fi
}
\newcommand{\nnnlo}{
  \ifmmode \textnormal{N3LO} \else\unskip N3LO\fi
}

\begin{document}
\title{Inference of the low-energy constants in $\Delta$-full chiral
  effective field theory including a correlated truncation error}
\date{\today}

\author{Isak Svensson}
\affiliation{Department of Physics, Chalmers University of Technology, SE-412 96 G\"oteborg, Sweden}

\author{Andreas Ekstr\"om}
\affiliation{Department of Physics, Chalmers University of Technology, SE-412 96 G\"oteborg, Sweden}

\author{Christian Forss\'en}
\affiliation{Department of Physics, Chalmers University of Technology, SE-412 96 G\"oteborg, Sweden}

\begin{abstract}
We sample the posterior probability distributions of the low-energy constants (LECs) in $\Delta$-full chiral effective field
theory (\chieft) up to third order. We use eigenvector
continuation for fast and accurate emulation of the likelihood and Hamiltonian Monte Carlo to draw effectively independent samples from the posteriors. Our Bayesian inference is conditioned on
the Granada database of neutron-proton (\np) cross sections and polarizations. We use
priors grounded in \chieft\ assumptions and a Roy-Steiner analysis of
pion-nucleon scattering data.
We model correlated EFT truncation errors using a two-feature Gaussian
process, and find correlation lengths for \np\ scattering energies and
angles in the ranges 45--83 MeV and 24--39 degrees, respectively. These
correlations yield a non-diagonal covariance matrix and reduce the
number of independent scattering data with a factor of 8 and 4
at the second and third chiral orders, respectively.
The relatively small difference between the second and third order
predictions in $\Delta$-full \chieft\ suppresses the marginal variance of
the truncation error and the effects of its correlation structure. Our results are particularly important for analyzing the predictive capabilities in \textit{ab initio} nuclear theory.
\end{abstract}

\maketitle

\section{Introduction}
\label{sec:intro}
A chiral effective field theory (\chieft) description of the nuclear
interaction~\cite{Weinberg:1990rz,Weinberg:1991um,Epelbaum:2008ga,Machleidt:2011zz,Hammer:2019poc}
is endowed with a power counting (PC) to organize the order-by-order
contributions of the strong-interaction dynamics to nuclear
observables. This kind of organization, regardless of the particulars
of the adopted PC~\cite{Yang:2019hkn}, is a hallmark of
EFT~\cite{Weinberg:1978kz} and \textit{ab
  initio}~\cite{Ekstrom:2022yea} approaches to nuclear theory since it
promises a handle on the theoretical uncertainty coming from
truncating the EFT expansion~\cite{Furnstahl:2015rha}. Accounting for
the truncation error is key to mitigating overfitting of the low-energy
constants (LECs) as well as assessing the importance of
discrepancies. Indeed, the modeling of EFT truncation errors can play
a significant role in the robustness of LEC inferences and
ensuing nuclear
predictions~\cite{Ekstrom:2015rta,Epelbaum:2014efa,Carlsson:2015vda,Stroberg:2019bch,Wesolowski:2021cni,Jiang:2022tzf,Jiang:2022oba}. \citet{Melendez:2019izc}
have proposed a Bayesian model for the truncation error that accounts for
finite correlations across independent variables, e.g., the scattering
energy and angle for nucleon-nucleon (\NN) scattering cross sections
and polarizations. To date, LEC inference in \chieft\ typically
accounts for truncation errors in the fully correlated or uncorrelated
limits~\cite{Melendez:2017phj,Wesolowski:2015fqa,Wesolowski:2018lzj,Svensson:2021lzs,Svensson:2022kkj},
and the robustness with respect to the correlation structure is not
well known.

In this paper, we quantify a correlated truncation error and analyze
its effects on a Bayesian estimation of the LEC posteriors for a
$\Delta$-full \chieft\ description of the neutron-proton (\np)
interaction up to next-to-next-to-leading order
(\nnlo)~\cite{Krebs:2007rh,Ekstrom:2017koy}. This extends our previous
work on Bayesian LEC estimation in $\Delta$-less \chieft\ where we
employed an uncorrelated truncation
error~\cite{Wesolowski:2021cni,Svensson:2021lzs,Svensson:2022kkj}. We also use eigenvector continuation (EC)~\cite{Konig:2019adq},
i.e., a reduced basis method~\cite{Drischler:2022ipa}, to
efficiently and accurately emulate the scattering amplitudes entering
the \np\ scattering-data likelihood. Following our previous
publications~\cite{Svensson:2021lzs,Svensson:2022kkj}, we employ
Hamiltonian Monte Carlo (HMC)~\cite{duane87} to draw effectively independent samples
from the LEC posteriors.

The $\Delta(1232)$ resonance plays an important role in nuclear
physics since it represents a rather low excitation energy and couples
strongly to the pion-nucleon ($\pi N$) system. This was recognized
already in early \chieft\ descriptions of the
\NN\ interaction~\cite{Ordonez:1993tn,vanKolck:1994yi,Ordonez:1995rz},
and several modern \chieft\ interactions incorporate the $\Delta$ as
well~\cite{Piarulli:2014bda,Piarulli:2016vel,Logoteta:2016nzc,Jiang:2020the}.
In $\Delta$-full \chieft\, there are four subleading $\pi N$ LECs up
to \nnlo, usually denoted $c_1,c_2,c_3,c_4$. They govern the strength
of subleading $2 \pi$-exchange diagrams of the \NN\ interaction and
the leading three-nucleon (\NNN) $2\pi$ exchange with an intermediate
$\Delta$ excitation, the so-called Fujita-Miyazawa force
~\cite{Fujita:1957zz,Epelbaum:2007sq}.
A Roy-Steiner analysis of $\pi N$ scattering amplitudes by
\citet{Hoferichter:2015hva} has enabled a determination
of the $\pi N$ LECs. Unfortunately, the relatively unknown value of
the axial $\pi N \Delta$ coupling $h_A$ propagates to approximately
five times greater uncertainties for $c_2,c_3,c_4$ (compared to a
determination in $\Delta$-less \chieft{}) when matching
$\Delta$-full \chieft\ to the subthreshold amplitudes in the
Roy-Steiner formalism~\cite{Siemens:2016jwj}. A well-founded
truncation error would therefore pave the way for learning more about
the strength of subleading 2$\pi$ exchange, and leading \NNN\ forces,
from \NN\ scattering data.

This paper is organized as follows: In Sec.~\ref{sec:exp_theo} we
present our statistical model for linking experiment and \chieft. In
Sec.~\ref{sec:inference} we discuss our priors and likelihood, and in
Sec.~\ref{sec:correlated_error} we introduce the two-feature Gaussian-process model
of the correlated EFT truncation error. In Sec.~\ref{sec:emulation} we
discuss the training of EC \np\ scattering emulators. In
Sec.~\ref{sec:sampling} we present the results from HMC sampling of
the LEC posteriors. A summary and an outlook are given in
Sec.~\ref{sec:summary}.

\section{Linking experiment and theory}
\label{sec:exp_theo}
Following our previous
papers~\cite{Wesolowski:2021cni,Svensson:2021lzs,Svensson:2022kkj}, we
relate an experimental measurement $\yexp$ of some scattering
observable with a theoretical prediction $\yth^{(k)}$, up to chiral
order $k$, using an additive model to account for the respective
uncertainties, $\dyexp$ and $\dyth^{(k)}$:
\begin{equation}
  \yexp(\idpt) = \yth^{(k)}(\lecs ; \idpt) + \dyexp(\idpt) + \dyth^{(k)}(\idpt).
  \label{eq:exp_theo}
\end{equation}
The theory prediction $\yth$ depends deterministically on the vector
of LECs $\lecs$ and the independent variable $\idpt = (\tlab,\theta)$,
where $\tlab$ denotes the kinetic energy of the incoming nucleon in
the laboratory frame and $\theta$ denotes the scattering angle in the
center-of-mass frame. For the total scattering cross section we have
$\idpt = (\tlab)$ as this observable is integrated over all
$\theta$. We have suppressed the explicit $\lecs$-dependence of
$\dyth^{(k)}$ as we will assume a fixed parameter value in the model
of the theory error.

The composition of the LEC vector $\lecs$ depends on the chiral order.
For the \np\ potentials in this paper, the LEC vector up to each order
is given by
\begin{equation}
\lecs_{\lo} = \Bigl(\Ct_{1S0}, \Ct_{3S1}\Bigr),
\end{equation}
\begin{align}
\begin{split}
\lecs_{\nlo} &= \Bigl(\Ct_{1S0}^{np}, \Ct_{3S1}, C_{1S0}, C_{3P0}, \\
&C_{1P1}, C_{3P1}, C_{3S1}, C_{3S1-3D1}, C_{3P2}\Bigr),
\end{split}
\end{align}
\begin{align}
\begin{split}
\lecs_{\nnlo} &= \Bigl(\Ct_{1S0}^{np}, \Ct_{3S1}, C_{1S0}, C_{3P0}, C_{1P1}, C_{3P1}, \\ & C_{3S1}, C_{3S1-3D1}, C_{3P2}, c_1, c_2, c_3, c_4 \Bigr),
\end{split}
\end{align}
where \lo\ is leading order ($k=0$), \nlo\ is next-to-leading-order
($k=2$), and \nnlo\ is next-to-next-to-leading order
($k=3$).\footnote{Due to
symmetries~\cite{Epelbaum:2008ga,Machleidt:2011zz,Hammer:2019poc} the
order $k=1$ vanishes.} We employ units and a notation linked to the
momentum partial-wave basis; see
Refs.~\cite{Machleidt:2011zz,Krebs:2007rh} for details. The potential
is nonlocally regulated with a regulator cutoff $\Lambda=450$ MeV as
in Ref.~\cite{Ekstrom:2017koy}. In the following we will only refer to
the generic vector $\lecs$, while the specific chiral order, if
important, should be obvious from the context.

The power counting of $\Delta$-full \chieft\ allows us to express a
prediction $\yth^{(k)}(\lecs;\idpt)$ up to chiral order $k$ as a sum
of order-by-order contributions
\begin{equation}
  \label{eq:eft_expansion}
  \yth^{(k)}(\lecs;\idpt) =  \yref(\idpt)\sum_{i=0}^{k}c^{(i)}(\lecs;\idpt)Q^i,
\end{equation}
where $\yref$ is some characteristic scale to be decided, and
$c^{(i)}$ are dimensionless expansion coefficients. The dimensionless
expansion parameter $Q=f(m_{\pi},\delta,p)/\Lambda_{\chi}$ is a ratio
composed of a soft scale given by some function of the pion mass
($m_{\pi}$), the $\Delta N$ mass splitting $\delta$, the (external)
relative \NN\ momentum $p$, and a hard scale $\Lambda_{\chi}$ for
which we assume a point-estimate value of 600
MeV~\cite{Reinert:2017usi,Melendez:2017phj}. We treat $\delta \sim
m_{\pi}$ as a small scale~\cite{Hemmert:1997ye}, although it does not
vanish in the chiral limit. We resum the potential to all orders
which obfuscates the form of $f$, and in line
with~\cite{Melendez:2019izc} we therefore assume the following
functional form
\begin{equation}
  f(m_{\pi},p) = \frac{p^8 + m_\pi^8}{p^7 + m_\pi^7},
\end{equation}
which facilitates a smooth transition across the soft scale
$m_{\pi}$. We find that the exact form of this function does not
impact our inference significantly, and reverting to a canonical
estimate $f = \text{max}(m_{\pi},p)$ does not change any of our
results. The upshot of having an order-by-order description of $\yth$,
as in Eq.~\eqref{eq:eft_expansion}, is a handle on the theoretical
uncertainty via
\begin{equation}
  \dyth^{(k)}(\idpt) = \yref \sum_{i=k+1}^\infty c^{(i)}(\idpt)Q^i.
  \label{eq:eft_error}
\end{equation}
Clearly, we have not explicitly extracted EFT coefficients $c^{(i>k)}$ for any
$\idpt$ and as a consequence we are uncertain about their
values. However, if we assume naturalness to hold, it is reasonable to
have $c^{(i)}\sim \mathcal{O}(1)$ for \textit{all} expansion
coefficients, including the unseen ones.
Furthermore, contributions can be either positive or negative.
We will therefore consider
the coefficients as drawn from some underlying probability distribution bounded by
a finite variance $\cbarsq$ expected to be of order
1. In addition, there is most likely information about $\yth(\idpt)$
at nearby values of the independent variable. As such, there is likely
nonzero covariance between expansion coefficients $c^{(i)}( \idpt_m)$
and $c^{(i)}( \idpt_n )$ at different values $\idpt_m$ and
$\idpt_n$. That correlation structure is of primary interest in this
paper.

\section{Setting up the LEC parameter inference}
\label{sec:inference}
We seek a posterior probability density function (\pdf)
$\prob(\lecs|\data,I)$ for the LECs $\lecs$ in $\Delta$-full
\chieft\ up to \nnlo\ conditioned on selected \np\ scattering data
$\data$ from the Granada database~\cite{perez13-1,perez13-2} and other
information $I$ as specified below. In the following, we use bold
symbols to denote quantities that depend on a range of values for the
independent variable $\idpt$, e.g., $\peth = (\idpt_1,\idpt_2,\ldots)$
and $\data = (\yexp(\idpt_1),\yexp(\idpt_2),\ldots)$, where we let $y$
denote observable type (differential cross section, vector
polarization, etc.).

Using Bayes's rule we can express the posterior in terms of a
likelihood $\prob(\data|\lecs,I)$, prior $\prob(\lecs|I)$, and
evidence $\prob(\data|I)$ as
\begin{equation}
  \prob(\lecs|\data,I) = \frac{\prob(\data|\lecs,I)\prob(\lecs|I)}{\prob(\data|I)}.
\end{equation}
We ignore the evidence term in this paper as it provides overall
normalization and does not impact the shape of the posterior.
\subsection{Prior}
\label{sec:prior}
We assume independent priors for the LECs
of the \NN\ contact and $\pi N$ potentials, i.e.,
\begin{equation}
  \prob(\lecs|I) = \prob(\nnlecs|I) \times \prob(\pinlecs|I),
\end{equation}
where $\pinlecs = (c_1,c_2,c_3,c_4)$\footnote{Not to be confused with
the EFT expansion coefficients $c^{(0)},c^{(2)},\ldots$ in
Eq.~\eqref{eq:eft_expansion}.} denotes the subleading $\pi N$ LECs. For
these, we place a normal and correlated prior based on the
(nondiagonal) covariance matrix $\cov_{\pi N}$
obtained in the Roy-Steiner analysis of \piN\ scattering data by \citet{Siemens:2016jwj},
\begin{equation}
  \prob(\pinlecs|I) = \mathcal{N}(\vec{\mu}_{\pi N},\cov_{\pi N}),
  \label{eq:rs_prior}
\end{equation}
with
\begin{equation}
  \cov_{\pi N} = \left[
    \begin{array}{cccc}
+6.11 & -0.63   & +6.26   & +0.25 \\
-0.63 & +277.43 & -359.84 & +174.20 \\
+6.26 & -359.84 & +474.01 & -224.92 \\
+0.25 & +174.20 & -224.92 & +119.16
    \end{array}
    \right]\times \frac{10^{-4}}{\text{GeV}^2}
\end{equation}
and
\begin{equation}
  \vec{\mu}_{\pi N} = [-0.74,-0.49,-0.65,+0.96] \cunit.
\end{equation}
Along the lines of naturalness, we place a normal and uncorrelated
prior on the LECs $\nnlecs$ of the contact potential,
\begin{equation}
  \prob(\nnlecs|I) = \mathcal{N}(\vec{0},\cov_{NN}),
\end{equation}
where $\cov_{NN}$ is a diagonal covariance matrix with standard
deviations $5 \times 10^4$ GeV$^{-(k+2)}$ for the contact LECs first
appearing at orders $k = 0, 2$. This is the same prior as in our
previous papers~\cite{Svensson:2021lzs,Svensson:2022kkj}.

\subsection{Likelihood}
\label{sec:likelihood}
\begin{figure*}[htpb]
\includegraphics{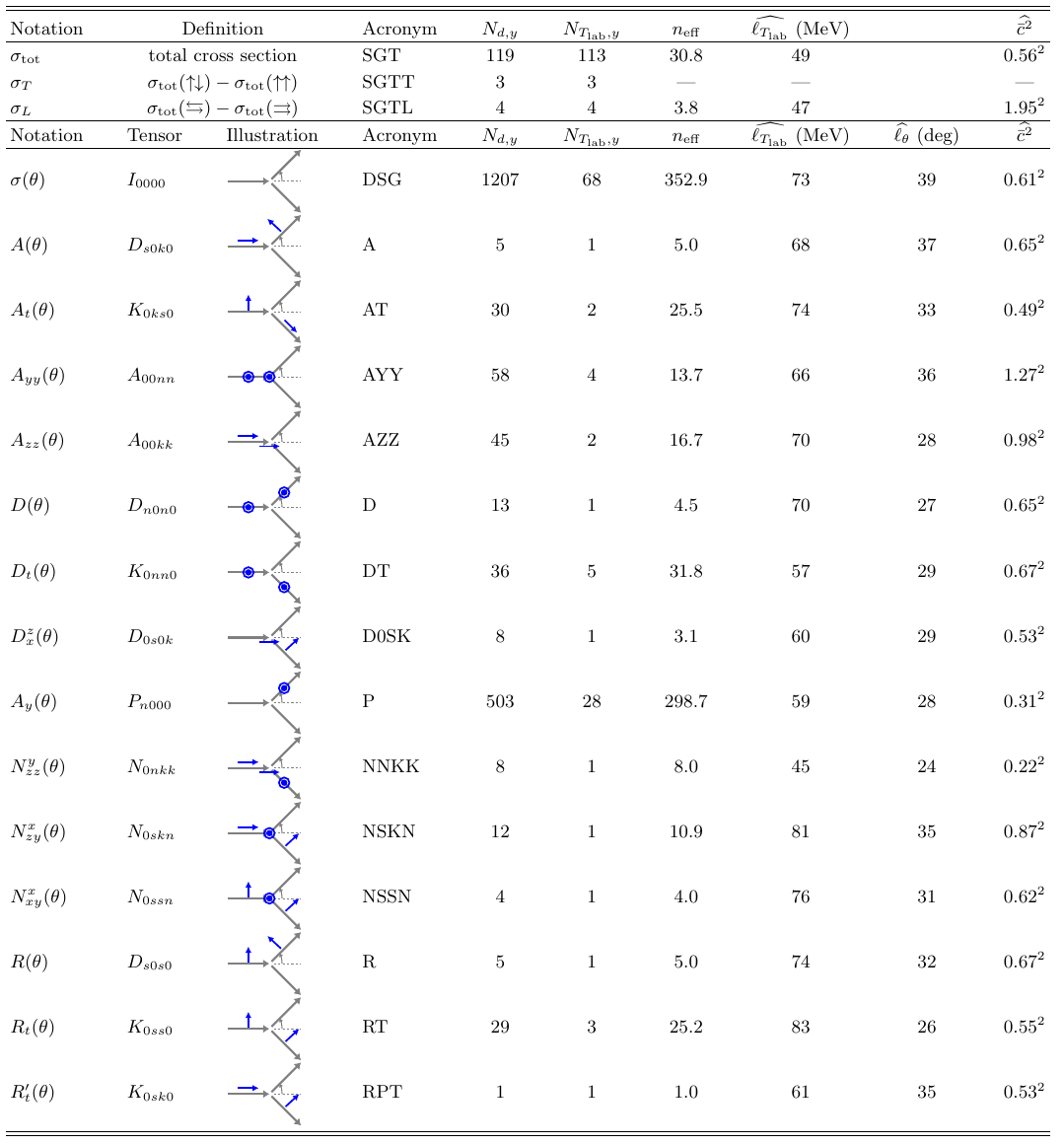}
\caption{The database $\data$ of \np\ scattering observables with
  $30\leq\tlab\leq290$ MeV with notation, definition, and acronym in the Granada
  database~\cite{perez13-1,perez13-2}, number of data $\ndatay$ and
  measurement energies $\nEy$ per observable type $y$, effective
  dimension $n_\text{eff}$ given theory error correlations, maximum
  \textit{a posteriori} (MAP) values for $\GP$ correlation lengths across
  scattering energy $\widehat{\lsE}$ and angle $\widehat{\lsth}$, and
  marginal variance $\widehat{\cbarsq}$. The spin-dependent
  integrated cross sections $\sigma_{T}$ and $\sigma_L$ have
  polarizations parallel and anti-parallel to each other and
  transverse ($T$) or longitudinal ($L$) to the beam direction. The
  $\sigma_T$ observable was omitted in our analysis, see the text for
  details. For spin polarizations we list the polarization tensors, $X_{srbt}$,
  using the notation of~\cite{Bystricky:1976jr} where
  subscripts denote the polarization of the (s)catter, (r)ecoil,
  (b)eam, and (t)arget particles in the direction of the incoming
  projectile $(\hat{k})$, normal to the scattering plane $(\hat{n})$,
  $\hat{s} = \hat{n}\times \hat{k}$, and 0 is an unpolarized
  state. The illustrations show initial and final spin-polarizations
  in the laboratory frame using gray arrows for reactants, blue arrows
  for the polarization direction in the scattering plane, and
  circled dots for the outward-pointing polarization vector. The projectile
  impinges from the left, the target (not indicated) is located in the
  center, and the projectile (recoil) scatters upwards (downwards) an angle
  indicated by the arc (gray). In this table, and in the text, we leave the energy dependence of all observables implicit.}
\label{tab:database}
\end{figure*}

We consider only strong interaction effects in elastic reaction channels, and neglect all
electromagnetic interactions in our calculations. Thus, we condition
our LEC inferences on low-energy \np\ scattering data below the
pion-production threshold $\tlab=290$ MeV, and omit \np\ scattering
data with $\tlab < 30$ MeV as the nonzero magnetic moments of
neutrons and protons can distort the low-energy \np\ scattering
amplitude~\cite{Stoks:1990us,Machleidt:2000ge}. We partition the
experimental data set as $\data = \{\data_y\}_{y=1}^{\nobs}$, where
$\nobs=18$ is the number of unique types of scattering cross sections
and polarizations in the considered energy range. In Fig.~\ref{tab:database} we summarize in tabular form the data
of measured \np\ scattering observables that we include, as well as
some of the results that will be discussed below.

We assume that experimental and theoretical errors are
independent of each other such that the total covariance matrix can be
written as
\begin{equation}
  \covmtx = \covmtx_\text{exp} +  \covmtx_\text{th}.
\end{equation}
The (diagonal) covariance matrix
$\covmtx_{\text{exp}}$ is provided by the experimenters and we employ
normalization factors from the Granada database.  We model the
covariance of the EFT errors independently for each type of
\np\ scattering observable $y$, and make specific assumptions and
choices $I_y$ per observable type. The (block-diagonal) covariance
matrix $\covmtx_\text{th}$ is given by
\begin{equation}
  \covmtx_\text{th} = \left[
    \begin{array}{cccc}
      \covmtx_{\text{th},1} &  \boldsymbol{0} & \cdots & \boldsymbol{0}\\
      \boldsymbol{0} & \covmtx_{\text{th},2} & \cdots & \boldsymbol{0}\\
      \vdots         & \vdots                          & \ddots & \vdots \\
      \boldsymbol{0} & \boldsymbol{0} & \cdots & \covmtx_{\text{th},\nobs}
    \end{array}
    \right],
  \label{eq:full_covariance_matrix}
\end{equation}
where the (non-diagonal) covariance matrix $\covmtx_{\text{th},y}$ at
chiral order $k$ contains elements
\begin{equation}
  \label{eq:covariance_elements}
(\covmtx_{\text{th},y})_{mn} =
\text{cov}[\dyth^{(k)}(\idpt_m),\dyth^{(k)}(\idpt_n)]
\end{equation}
that we model using a Gaussian process; see
Sec.~\ref{sec:correlated_error}.

In accordance with the principle of maximum entropy we employ a
normally distributed data likelihood, which factorizes as
\begin{equation}
  \prob(\data|\lecs,I) = \prod_{y=1}^{\nobs} \prob(\data_y|\lecs,I_y),
  \label{eq:likelihood}
\end{equation}
 where
\begin{equation}
  \prob(\data_y|\lecs,I_y) \propto \exp[ -\frac{\boldsymbol{r}_y^T(\lecs)(\covmtx_{\text{exp},y} + \covmtx_{\text{th},y})^{-1}\boldsymbol{r}_y(\lecs)}{2}].
\end{equation}
Here, $\boldsymbol{r}_y(\lecs) =
[r_{y,1}(\lecs),r_{y,2}(\lecs),\ldots,r_{y,\ndatay}(\lecs)]$ is a
(column) vector of $\ndatay$ residuals, each one given by the
difference between experimental and order-$k$ theoretical prediction
values for the independent variable $\idpt$, i.e.,
\begin{equation}
  r_{y,j}(\lecs) = [y_{\text{exp}} (\idpt_j ) - y_{\text{th}}^{(k)} (\lecs;\idpt_j)].
\label{eq:residual}
\end{equation}
All cross sections are computed from amplitudes obtained numerically
via the nonrelativistic Lippmann-Schwinger
equation~\cite{Svensson:2021lzs,Carlsson:2015vda}. This approach is
virtually exact and we do not account for any method uncertainties of
numerical origin in this paper.
Note that in Eq.~\eqref{eq:residual} we assume that all errors taken into
account have mean zero, which will be the case in this paper.

\section{Correlating the EFT truncation error in \NN\ scattering}
\label{sec:correlated_error}
Following the suggestion by~\citet{Melendez:2019izc}, we model the EFT
truncation error using Gaussian processes
($\GP$s)~\cite{rasmussen2005gaussian,bishop2007}. This results in a covariance
matrix $\covmtx_{\text{th}}$ from Eq.~\eqref{eq:full_covariance_matrix} with
off-diagonal elements.
In this section, we
limit our discussion to the $\GP$ modeling of a single covariance
submatrix $\covmtx_{\text{th},y}$ for a specific observable type $y$,
e.g., the differential cross section $\sigma(\theta)$, and therefore
omit the observable index $y$ in the following.

We develop a $\GP$ model to handle two features, i.e.,
$\GP: \mathbb{R}^{2} \rightarrow \mathbb{R}^1$, for correlating the
distribution of expansion coefficients at different values of the
scattering energy \textit{and} angle. For total cross sections we only
operate with the scattering energy, as all angles are integrated
over. We expect unseen coefficients $c^{(i)}(\peth) \equiv
\{c^{(i)}(\idpt_j)\}_j$ to be distributed as
\begin{equation}
  c^{(i)}(\peth) \given m, \ls, \cbarsq \sim \GP[m, \cbarsq \ckernel(\peth',\peth;\ls)],
  \label{eq:gp_c}
\end{equation}
where $m$ is assumed to be a constant mean function and $\ls,\cbarsq$
denote the hyperparameters for the $\GP$ correlation length(s) and
marginal variance, respectively. The calibration of these parameters will be discussed
in Sec.~\ref{sec:calibration}. The marginal standard deviation $\cbar$ is a scale factor
that quantifies the average distance of the $\GP$ away from its mean,
and the correlation length quantifies the typical extent of the $\GP$
fluctuations. Elastic \np\ scattering observables typically exhibit a
smooth behavior as a function of energy and angle. Therefore, we
expect smooth variations of the expansion coefficients. Assuming stationarity, we
employ a squared-exponential kernel to parametrize the correlation
structure,
\begin{equation}
  \ckernel(\idpt_i,\idpt_j;\ls)  = \exp\left[ - \frac{(\idpt_i-\idpt_j)^T \ls^{-1}(\idpt_i-\idpt_j)}{2}\right],
\end{equation}
where the length scale(s) are given by
\begin{equation}
  \ls =\left[
    \begin{array}{cc}
      \lsE^2 & 0 \\
      0     & \lsth^2
  \end{array}\right],
\end{equation}
in an obvious notation.

We are mainly interested in the quantified hyperparameters $\ls$
and $\cbarsq$ that characterize the finite correlation length and variance
of the EFT error.  A $\GP$ is a linear-Gaussian model for which
there are closed-form
expressions~\cite{rasmussen2005gaussian,bishop2007} for the
distribution of predicted data conditioned on calibration data. Once
the posterior distribution, or point estimates, of the hyperparameters
are determined it is straightforward to quantitatively use the $\GP$
to predict EFT expansion coefficients exhibiting a variance and
correlation consistent with the calibration data.  Furthermore, one
can show~\cite{Melendez:2019izc} that the $\GP$ for the EFT truncation
error pertaining to a scattering observable $\yth$ is given by
\begin{equation}
\dyth^{(k)} (\peth) \given m,\ls,\cbarsq, \yref,Q \sim \GP[M^{(k)}(\peth),\cbarsq \ykernel^{(k)}(\peth',\peth;\ls)],
\end{equation}
where the respective mean ($M$) and kernel ($\ykernel$) functions are given by
\begin{equation}
  M^{(k)}(\peth) = \yref(\peth)\frac{[Q(\peth)]^{k+1}}{1-Q(\peth)}m,
\end{equation}
and
\begin{align}
  \begin{split}
    \ykernel^{(k)} (\peth',\peth;\ls) = {}& \yref(\peth')\yref(\peth) \\
    {}& \times \frac{[Q(\peth')Q(\peth)]^{k+1}}{1-Q(\peth')Q(\peth)}\ckernel(\peth',\peth;\ls).
  \end{split}
\end{align}
This lays the foundation for how we model the covariance matrix in
Eq.~\eqref{eq:full_covariance_matrix} and sample the LEC posterior
$\prob(\lecs|\data,I)$ via the likelihood and prior defined in
Secs.~\ref{sec:prior} and \ref{sec:likelihood}.
As will be discussed shortly, we set $m = 0$ in this paper.

\subsection{Optimizing the $\GP$ hyperparameters}
\label{sec:calibration}
We set up individual $\GP$s for the block-diagonal covariance matrices
$\covmtx_{\text{th},y}$ in Eq.~\eqref{eq:full_covariance_matrix}. The
hyperparameters for each observable type $y$ are optimized separately and we find
maximum \textit{a posteriori} (MAP) values for them using \texttt{gsum}~\cite{Melendez:2019izc}. For each $\GP$ we use a set of training data $\ctrain$
consisting of $N_{\text{train}}$ expansion coefficients obtained as
\begin{equation}
  c^{(k)}(\idpt_j) = \frac{\yth^{(k)}(\lecs;\idpt_j) - \yth^{(k-1)}(\lecs;\idpt_j)}{\yref [Q(\idpt_j)]^k},
  \label{eq:order_by_order}
\end{equation}
for a training range $\peth^{\star}$ of values for the independent
variable. We pick $N_{\text{train}} \approx$ 25--50 points uniformly
distributed across approximately 4--6 energies $\tlab \in[30,290]$ MeV
and 5--10 angles $\theta \in [0,180]$ degrees. The training grid
varies somewhat depending on the observable type. The order-by-order
differences in Eq.~\eqref{eq:order_by_order} are computed from
theoretical predictions based on a MAP estimate for the LECs,
$\lecs=\lecs^{\star}$, obtained in line with
Ref.~\cite{Svensson:2021lzs} in the uncorrelated limit using $\cbar =
1$.

\begin{figure}[bt]
\centering
\includegraphics[width=1.0\linewidth]{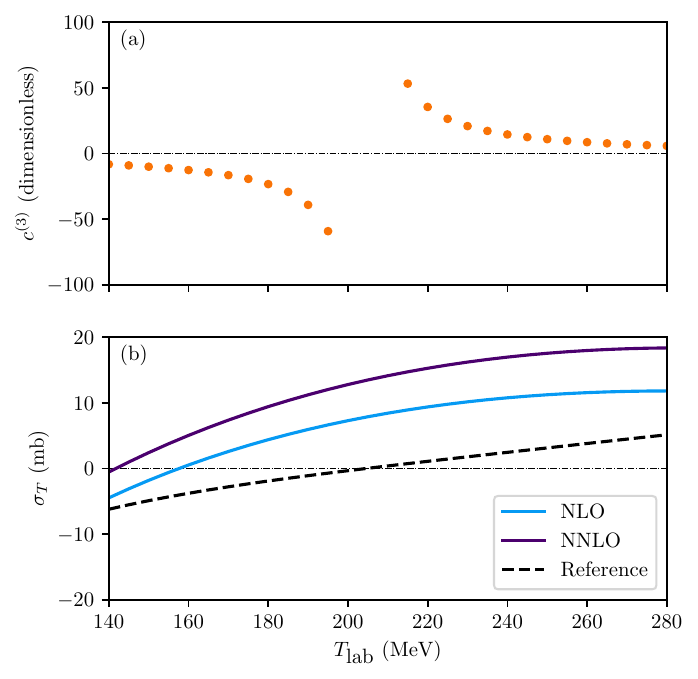}
\caption{(a) Calculated expansion coefficients for $\sigma_T$. (b) The \nlo,
\nnlo, and reference values from which the coefficients are calculated. Note the
zero crossing for the reference value at $\Tlab \approx 205$ MeV.}
\label{fig:c_SGTT}
\end{figure}

The LO predictions $\yth^{(0)}$ in $\Delta$-full \chieft\ are
identical to the $\Delta$-less predictions and as such they are rather
poor, as expected in Weinberg
PC~\cite{Carlsson:2015vda,Yang:2020pgi}. Thus, the $c^{(2)}$ expansion
coefficients incur rather unnatural values. One could argue that a
Bayesian approach, with a prior for the expansion parameters, should
be able to properly handle such outliers. However, a (likely)
deficient LO will conflict with our initial assumptions in
Eq.~\eqref{eq:eft_error}. Instead of using the truncation error to
absorb deficiencies rooted in the PC, we decided to neglect LO
predictions in this paper. Thus, all training data consist of $c^{(3)}$
expansion coefficients only. Despite this limited amount of data, we expect
the available coefficients to carry valuable information about the
correlated truncation error and we proceed with a Bayesian $\GP$ model.
For total and differential cross sections we set $\yref$ to the
predictions of the Idaho-N3LO(500) potential~\cite{Entem:2003ft},
which is almost the same as using $\yref = \yexp$. We encounter a
zero crossing in $\yref$ for $\sigma_T$. This leads to a discontinuity
in the $c^{(3)}$ coefficients as a function of $\tlab$; see
Fig.~\ref{fig:c_SGTT}. Such small-scale structures are virtually
impossible to faithfully represent using an infinitely smooth
squared-exponential kernel. As there are only three experimental data
points for this observable type, we decided to exclude them from
$\data$ rather than choosing an exceptional $\yref$ for this one case.
For polarization data we encounter an excess of small-scale
structures when setting $\yref$ to the predictions of
Idaho-N3LO(500). Using $\yref=0.5$ removes them entirely and we are
spared further kernel design. We choose 0.5 as a representative value
for these observables. Other choices could be $\yref=1.0$~\cite{Melendez:2017phj}
or an average of experimental values~\cite{Svensson:2021lzs}.

All training data are pruned for outliers using a three-interquartiles
threshold~\cite{Svensson:2021lzs}. In the end, all $c^{(3)}(\peth)$
coefficients are of natural order, and a vast majority of them pass
the outlier filter. Operating with a $\GP$ allows us to incorporate
known symmetry constraints on the polarization observables. Indeed,
some polarizations are identically equal to zero at $\theta=0$ and/or
$\theta=180$ degrees. We impose such boundary constraints by
adding zero-valued expansion coefficients for the endpoint(s) of the
angular direction to the training data. In a future application we
will formally incorporate continuous boundary constraints in the
$\GP$~\cite{vernon}.

\begin{figure}[t]
\centering
\includegraphics[width=0.99\columnwidth]{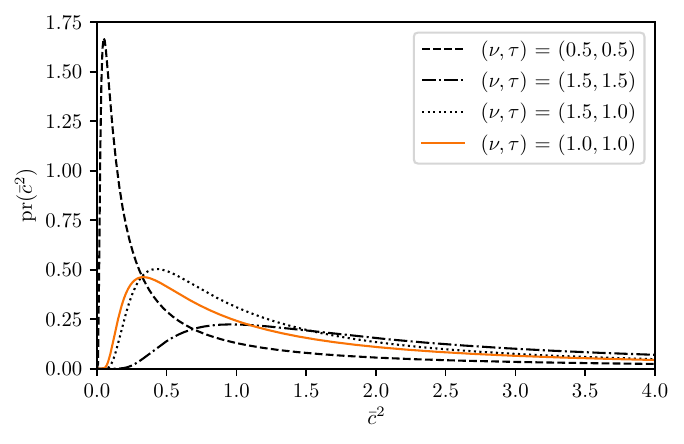}
\caption{ The inverse-$\chi^2$ prior $\prob(\cbarsq | I)$ (orange) for
  the $\GP$ hyperparameter $\cbarsq$. Shown in black are alternative
  priors that we use to show the robustness of the inference.}
\label{fig:conjugate_prior}
\end{figure}

We employ a bounded and uniform prior for the length scales $\ls$ of
the form
\begin{equation}
  \prob(\lsth|I) = \mathcal{U}(\epsilon,180)\,\,\text{degrees}
\end{equation}
and
\begin{equation}
  \prob(\lsE|I) = \mathcal{U}(\epsilon,290)\,\,\text{MeV},
\end{equation}
where $\epsilon = 10^{-5}$ is introduced to avoid numerical issues in
the optimization of the kernel (the exact value of $\epsilon$ is unimportant
in this case since the posterior values are never close to the edge of the prior).
We place a conjugate inverse-$\chi^2$ prior on the variance $\cbarsq$ according
to
\begin{equation}
  \prob(\cbarsq|I) = \chi^{-2}(\nu_0=1,\tau_0^2=1)
  \label{eq:mc2_prior}
\end{equation}
and a Dirac delta prior $\prob(m) = \delta(m)$ on the mean $m$.  This
encodes our expectation of natural $\cbarsq$ values, although the
heavy tail in our prior allows for some unnaturalness, and expansion
coefficients symmetrically distributed with mean $m \equiv 0$. Our
chosen prior for $\cbarsq$ is shown in orange in
Fig.~\ref{fig:conjugate_prior}.  We do not allow the mean to vary
owing to the limited amount of information---one chiral order---we
have available to learn from.  This strict prior on $m$ could be
substituted for, e.g., a Gaussian prior if more information is
available. If so, Eq.~\eqref{eq:residual} must be updated to include
the systematic contribution of a truncation error with non-zero mean.

\begin{figure*}[t]
\centering
\includegraphics[width=1.0\linewidth]{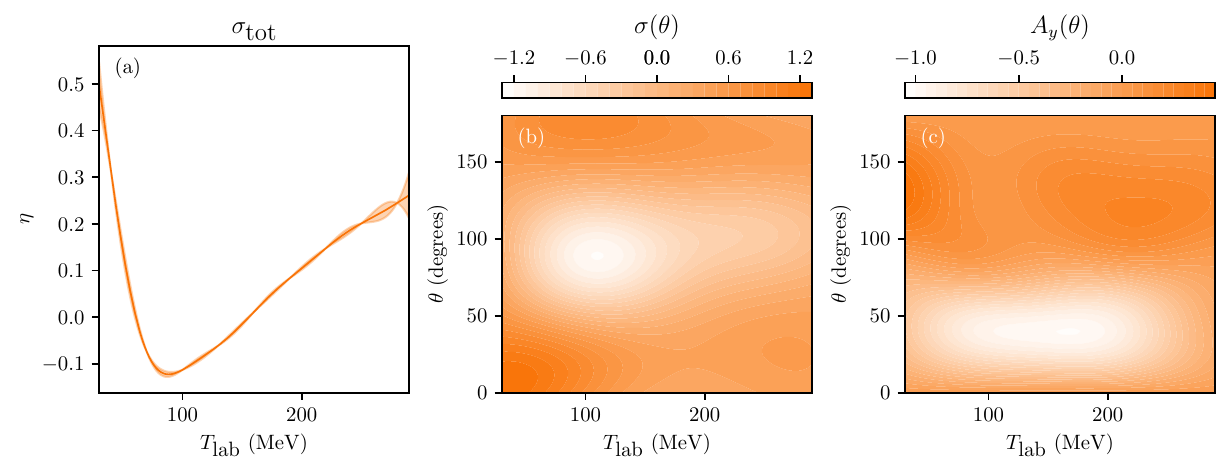}
\caption{Mean $\GP$ predictions $\eta$ for (dimensionless) expansion
  coefficients for (a) $\sigma_\text{tot}$ , (b)
  $\sigma(\theta)$, and (c) $A_y(\theta)$. The light colored
  band in (a) indicates a 95\% credible interval.}
\label{fig:gp_predictions}
\end{figure*}

The conjugacy of the $\chi^{-2}(\cbarsq)$ prior implies
that the posterior $\prob(\cbarsq|m=0, \ls,\ctrain,\peth,I)$ is
also an inverse-$\chi^2$ \pdf{} with updated
parameters~\cite{Melendez:2019izc}. The prior for the length scale
$\ls$ is not conjugate, and to find the posterior MAP values
$(\widehat{\ls},\widehat{\cbarsq})$ of
\begin{align}
  \begin{split}
    \prob(\ls,\cbarsq|m=0,\ctrain,\peth^{\star},I) \propto {}& \prob(\ctrain|m=0,\ls,\cbarsq,\peth^{\star},I) \\
    {}& \times \prob(\cbarsq|m=0,I)\prob(\ls|I)
  \end{split}
\end{align}
we use the numerical optimization routine
L-BFGS-B~\cite{byrd1995limited} with multiple restarts. We include a
so-called nugget, set to $10^{-10}$, to avoid numerical issues.  The
resulting MAP estimates of the hyperparameters are listed in
the table in Fig.~\ref{tab:database}. The delta prior on $m$ obviously yields a
corresponding MAP value $\widehat{m} = 0$. The $\GP$ predictions for
expansion coefficients $c^{(3)}$ for three of the most abundant
observable types are shown in Fig.~\ref{fig:gp_predictions}.

The amount of training data is sufficient to make the posterior
likelihood-dominated in most cases. We explore a number of priors, shown
in Fig.~\ref{fig:conjugate_prior}, and find that
the inference is rather robust as long as we allow for very small
values $\cbarsq \gtrsim 0.2^2$. Indeed, a prior that is too restrictive
in this regard can bias the result for total cross sections, where the
training data are one dimensional.

\subsection{Validating the $\GP$ hyperparameters}
\label{sec:validation}
We validate the $\GP$ hyperparameters for each observable type $y$
using a set of $\nval$ complementary validation data $\cval$. The
validation data are generated by the same method as the training data,
but with energies and angles shifted so that the validation values
$\xval$ for the independent variable do not overlap with ones used
during training.  In addition to visual inspections, which are
certainly useful, we also employ a set of diagnostics inspired by
Refs.~\cite{Bastos:2009,Melendez:2019izc}. These
diagnostics are the Mahalanobis distance, pivoted Cholesky errors, and
the credible interval diagnostic.  These, and a few more, are
thoroughly discussed in Ref.~\cite{Melendez:2019izc}.

The (squared) Mahalanobis distance, $\mdsquared$, is a multivariate analog to the sum of squared residuals. Here, it is computed as
\begin{equation}
  \mdsquared (\cval;\widehat{m}, \gpcovval) = [\cval -
    \widehat{m}]^T\gpcovval^{-1}[\cval - \widehat{m}],
\end{equation}
where $\widehat{m}$ is the $\GP$ mean at $\xval$---equivalent to zero with our choice of prior---and
\begin{equation}
  \gpcovval = \widehat{\cbarsq} \ckernel(\xval',\xval;\widehat{\ls}).
  \label{eq:gpcovval}
\end{equation}
This distance is
commonly used to quantify the deviation of a prediction compared to
data in a correlated setting. Here, we use it to diagnose whether a
set of validation data could reasonably have been drawn from a $\GP$ with covariance according to Eq.~\eqref{eq:gpcovval} and mean $m = \widehat{m} = 0$.
Either too large or small values of $\mdsquared$,
compared to the reference $\chi^2$ distribution with $\nval$ degrees
of freedom, point to a possible conflict between the $\GP$
and the validation data. The Mahalanobis distance is a scalar measure
and does not provide detailed insight about the possible tension with
respect to the validation data.

Furthermore, it is instructive to study the $\GP$ and the validation data point by point
as a function of the independent variable. Such comparisons are not straightforward
to interpret given the mutual correlation of the
validation data. Therefore, we decorrelate and rescale the covariance matrix
to independent unit variances using a Cholesky decomposition; $\gpcovval =
\boldsymbol{G}\boldsymbol{G}^T$ where $\boldsymbol{G}$ is a triangular
standard deviation matrix. From this we define the Cholesky errors
\begin{equation}
  \boldsymbol{D}_G = \boldsymbol{G}^{-1}[\cval - \widehat{m}].
\end{equation}
To order the vector $\boldsymbol{D}_G$ in a meaningful way we pivot
the decomposition in decreasing conditional
variances. One should not detect any pattern when
plotting the pivoted $\boldsymbol{D}_G$ versus the index of the
validation data. To reveal further information one can introduce a
ratio scale also on the abscissa by plotting the pivoted Cholesky
errors versus the conditional variances used for
pivoting~\cite{taweel18}.

All but one of the $\GP$s readily pass our diagnostics, with
$\mdsquared$ landing within 95\% credible intervals of the respective
reference distributions. We see little to no structure in the pivoted
Cholesky decompositions, and empirical coverages that roughly match
the corresponding credible intervals. In addition, visual inspections
indicate that the inferred hyperparameters are plausible. Naturally, incorporating expansion coefficients from other orders would provide a stronger test of our $\GP$ model. We only
reject $\sigma_T$, and therefore also the three experimental data
points. Neither $\mdsquared$ nor the $\boldsymbol{D}_G$ exhibit
significantly poor performance for this observable given our
validation data. However, upon visual inspection of the $c^{(3)}$
curve for this observable, shown in Fig.~\ref{fig:c_SGTT}, we find
that it is discontinuous for $\Tlab \approx 205$ MeV because the
reference value crosses zero at this point, leading to extremely large
coefficients near the crossing point and an abrupt change in
sign. Clearly, our $\GP$ model, based on a squared-exponential kernel,
is ill suited to handling this discontinuity, and the inferred length
scale is strongly dependent on the chosen training data and will
approach zero as we increase the number of training data. If it were
not for the existence of an experimental data point in the vicinity of
the problematic scattering energy we would have considered excluding
only the problematic region. This example clearly shows that one
should not blindly trust the diagnostics as they are conditioned on
the chosen training and validation data.

\subsection{The structure of the correlated truncation error}
Introducing a correlated truncation error reduces the number of
independent directions in the data space. Indeed, two data residing
within one correlation length of the scattering energy or angle
carry joint information. To quantify the impact of this we compute
the effective dimension $n_\text{eff}$, per observable type $y$, using a
measure~\cite{Hill1973,giudice21} defined as
$n_\text{eff} = \prod_{i=1}^{\ndatay} \normeig_i^{-\normeig_i}$,
where $\normeig_i$ denotes a normalized eigenvalue $\normeig_i =
\lambda_i/\text{tr}(\boldsymbol{C}_{y})$ of the respective correlation matrix. Here, $\boldsymbol{C}_y =
\boldsymbol{S}_y^{-1}\covmtx_y\boldsymbol{S}_y^{-1}$ and
$\boldsymbol{S}_y = \sqrt{\text{diag}(\covmtx_y)}$ with $\covmtx_y =
\covmtx_{\text{exp},y} + \covmtx_{\text{th},y}$. We could equally well
compute $n_\text{eff}$ for $\covmtx_y$ directly. However, operating with the
correlation matrix leads to the values $n_\text{eff}=1$ and
$n_\text{eff}=\ndatay$ in the limits of having a full off-diagonal
correlation and zero off-diagonal correlation,
respectively. Consequently, we interpret $n_\text{eff}$ as the
effective number of data, and note that the logarithm of $n_\text{eff}$
corresponds to the discrete Shannon entropy of the spectrum of the
normalized eigenvalues ${\normeig_i}$.

\begin{figure}[t]
\centering
\includegraphics[width=1.0\linewidth]{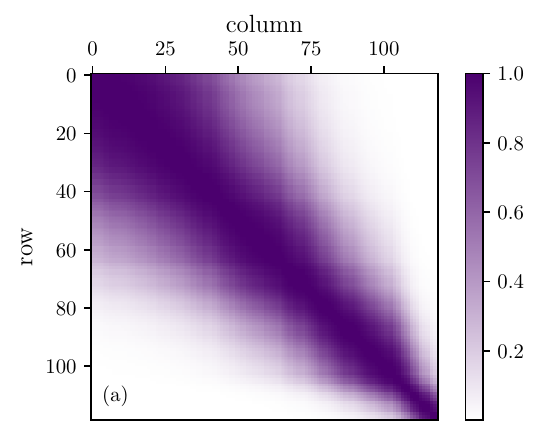}
\includegraphics[width=1.0\linewidth]{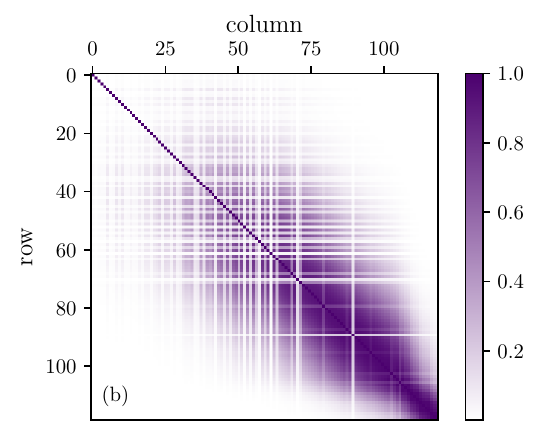}
\caption{Correlation matrices based on the \nnlo\ covariance matrix
  $\boldsymbol{\Sigma}_{\sigma_\text{tot}}$ of the 119
  $\sigma_\text{tot}$ data. In panels (a) and (b) we exclude and
  include, respectively, the experimental variances
  $\boldsymbol{\Sigma}_{\text{exp},\sigma_\text{tot}}$. The data are
  sorted by energy in increasing order.}
\label{fig:c_cov_SGT}
\end{figure}

The resulting $n_\text{eff}$ values at \nnlo\ are summarized in
the table in Fig.~\ref{tab:database}. They show that the correlations reduce
the number of effective data from 2087 to 841, which is still
plentiful. The main reason for the relatively weak impact of the
correlated truncation error is the dominance of the experimental
variances along the diagonal of the covariance matrix $\covmtx$.
Indeed, for the correlation matrices of the $\GP$ kernel for EFT
truncation errors alone we find $n_\text{eff} \approx 5-10$. In
Fig.~\ref{fig:c_cov_SGT} we show correlation matrices for the total
cross section $\boldsymbol{C}_{\sigma_\text{tot}}$, with and without
the experimental covariance matrix accounted for. The diagonal
dominance of the experimental errors is clearly visible. This
dominance weakens for higher energies (higher indices in the figure)
as the truncation error increases with the scattering energy $\tlab$.
At \nlo, where the truncation error is greater by one chiral order
$Q$, the corresponding $n_\text{eff}$ values are typically a factor
2 smaller as the correlations of the truncation error become
relatively more important. The total number of effective data is
498. In Fig.~\ref{fig:emulator_error}(a) we compare the 68\% credible interval $\Delta y_\text{th}^{(k)}$ of the EFT truncation error at
\nlo\ and \nnlo\ with the experimental
errors $\Delta y_\text{exp}$ for all data.

The EFT truncation error, and its correlations, is expected to be more
important for inferences conditioned on very precise \pp\ data. To estimate this
effect, we inspected the \pp\ $A_y(\theta)$ data set more closely. With
$\tlab \in [30,290]$ MeV there are 496 data points, which is almost
the same as in the \np\ sector. Reducing the experimental variances of
the \np\ $A_y(\theta)$ data, to mimic the average level of \pp\ variances, we
observe $n_\text{eff}=154$ and 91 at \nnlo\ and \nlo,
respectively. For other observable types, like $\sigma(\theta)$, the
average \pp\ variance is greater than the average \np\ variance, and
the opposite effect is likely observed. Note that the distribution of
\np\ and \pp\ measurement energies and angles differ and we do not
account for this in our estimate.

\section{Emulating $np$ scattering cross sections}
\label{sec:emulation}
We use an EC-based method~\cite{Konig:2019adq} to construct accurate
and efficient emulators for \np\ scattering observables. Operating
with emulators, instead of the exact simulators, helps reduce the
computational cost of evaluating the likelihood. As an added bonus, once the emulators are trained they are straightforward to distribute
in the scientific community.

\begin{figure}[t]
\centering
\includegraphics[width=1.0\linewidth]{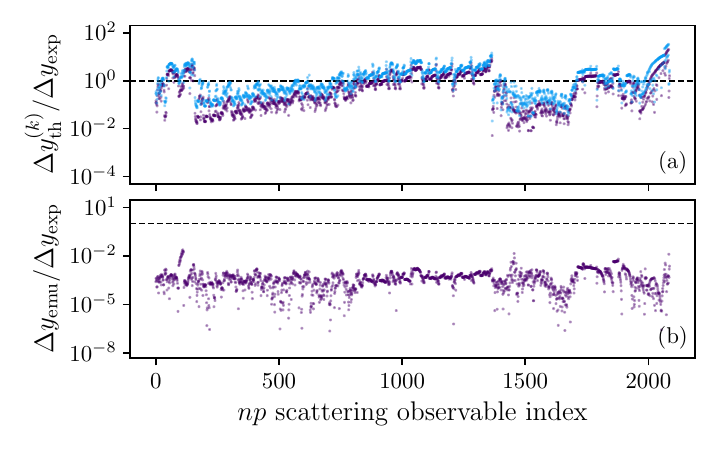}
\caption{The EFT truncation and EC emulator errors compared to
  experimental errors for all considered \np\ scattering data. (a) The
  standard deviations of the EFT truncation errors at \nlo\ and
  \nnlo. (b) The \nnlo\ EC emulator errors ($\Delta
  y_\textnormal{emu}$). Colors as in Fig.~\ref{fig:c_SGTT}. In both
  panels we divide by the the experimental $1\sigma$ errors ($\Delta
  y_\textnormal{exp}$) and the horizontal dashed lines indicate a
  ratio of 1.}
\label{fig:emulator_error}
\end{figure}

We use Newton's functional formulation of on-shell \np\ $T$-matrix
elements for setting up the
emulators~\cite{Melendez:2021lyq}. Technically, this leads to one
emulator per \np\ partial-wave and unique scattering energy in the
database $\data$ (table in Fig.~\ref{tab:database}). Truncating the
partial-wave expansion at a maximum \np\ angular-momentum quantum
number $J=30$ leads to 182 partial waves. With $\nE=177$ energies in
$\data$ we end up with $32\,214$ $T$-matrix emulators per chiral order.
The training values for the LECs are drawn according to a
latin hypercube design in a sufficiently wide interval $[-4,+4]$ (see
Sec.~\ref{sec:prior} for units). To simplify the setup, we employ the
same training protocol for all emulators and find it sufficient to use
7 (8) training points at \nlo\ (\nnlo). As there are at most $3$ ($7$)
relevant LECs acting per partial wave, this approach leads to very
accurate emulation of all relevant scattering observables. We estimate that emulator errors $\Delta
y_\text{emu}$ are at least 10, and typically 1000, times smaller than
experimental errors. This is quantified using the difference between emulated and
exact results for all \np\ observables in $\data$ at ten different
sets of LEC values following randomized latin hypercube designs within
the training intervals specified above. In Fig.~\ref{fig:emulator_error}(b) we show the ratio $\Delta y_\text{emu}/\Delta y_\text{exp}$ at \nnlo\ for a random set of LEC values. This is virtually identical to the \nlo\ result. We conclude that the emulator
errors are sufficiently small to neglect them in our inferences. In the following we
refer to the collection of all amplitude-emulators, per chiral order,
as the \textit{emulator} for scattering observables $y$.

\section{Sampling posterior distributions}
\label{sec:sampling}
We sample the LEC posteriors $\prob(\lecs|\data,I)$ at \nlo\ and
\nnlo\ using HMC, an advanced
Markov chain Monte Carlo (MCMC) algorithm suitable for high-dimensional sampling problems. We
have detailed the use and performance of HMC in a previous
paper~\cite{Svensson:2021lzs} and further improved its performance in
Ref.~\cite{Svensson:2022kkj}. Here, we employ the same sampling
strategy as in the latter paper. HMC needs tuning in order to perform
well, and in particular a good so-called mass matrix is needed. The
mass matrix should be a decent approximation to the inverse of the
parameter covariance matrix. We extract such an approximation by
maximizing the posterior through the standard BFGS
algorithm~\cite{broyden70,fletcher70,goldfarb70,shanno70}.

HMC
requires gradients of the (logarithm of the) posterior with respect to
the sampled parameters. The underlying linearity of the emulators furnishes easy access to derivatives of $T$-matrix amplitudes with respect to the LECs. However, we employ the automatic differentiation (AD) and just-in-time (JIT) compilation
tools for Python in the Google JAX~\cite{jax2018github} library to compute the required gradients and boost the execution speed
of our Python code; it takes on the order of one second to evaluate the entire data
likelihood, including derivatives. We opted for this approach due to its simplicity and speed.
It also enables straightforward computation of derivatives of the posterior
with respect to parameters other than the LECs, such as $\GP$ hyperparameters.
A threaded C-code implementation would likely
be more efficient, but not pivotal for the present applications.

We diagnose the convergence of the MCMC chains using a criterion
based on the integrated autocorrelation time
$\tau_\textnormal{int}$~\cite{Sokal}. This is a measure of the
autocorrelation between MCMC samples, and it tends to be
underestimated for short MCMC chains. In line with
Ref.~\cite{Svensson:2021lzs} we therefore declare convergence when (a)
the estimation of $\tau_\textnormal{int}$ has stabilized, and (b) when
$N \geq 50\tau_\textnormal{int}$, where $N$ is the length of the MCMC
chain. All our chains readily pass this test. Like other MCMC
convergence tests, the $\tau_\textnormal{int}$ criterion may falsely
declare convergence, e.g., if the posterior is multimodal; we
therefore search for local optima using BFGS optimization initialized
at random positions. We have not detected signs of multimodality and
are confident that our chains are converged.

\subsection{LEC posteriors}
In a first step we compare the LEC posteriors obtained with correlated and uncorrelated EFT truncation
errors. In the uncorrelated limit we determined $\cbarsq$ from the
root-mean-square value of the \nlo-\nnlo\ expansion coefficients for
the $\sigma_\text{tot},\sigma(\theta),A_y(\theta)$ observables, i.e.,
omitting the \lo\ contribution and discarding outliers. We employed
the same grids of scattering energies and angles as in the training of
the correlated $\GP$ model for these observables, see
Sec.~\ref{sec:calibration}. This leads to a fixed $\cbarsq=0.42^2$.

We find that the introduction of a finite correlation length in the EFT
truncation error does not affect the marginal LEC posteriors much; see
Figs.~\ref{fig:nlo_uncorrelated}, \ref{fig:nlo_correlated},
\ref{fig:nnlo_uncorrelated}, and \ref{fig:nnlo_correlated} in
Appendix~\ref{app:posteriors}. However, there are some differences and
similarities worth commenting on. Most noticeably, the marginal posteriors for all
LECs are approximately twice as wide when including a correlated
truncation error. This is consistent with the observed reduction in
the effective number of data.
Other than that, the posterior correlation structure
remains the same and the respective locations of the modes are largely
the same, except for a significant shift in the value of the
$C_{3S1-3D1}$ contact and the $C_{1P1}$ contact LEC becoming
(almost) consistent with zero in the correlated (uncorrelated) limit
at \nlo\ (\nnlo).
At \nnlo\, the MAP values for the subleading $\pi N$ LECs are
substantially shifted with respect to the mean of their prior values
in Eq.~\eqref{eq:rs_prior}. This result does not change when
neglecting a correlated truncation error. In the case of a
correlated truncation error we obtain a MAP value for the $\pi N$ LECs,
\begin{equation}
\widehat{\lecs}_{\pi N} = [-0.72(2),-0.96(5),-0.01(6),0.69(5)]\,\text{GeV}^{-1},
\end{equation}
with 68\% credible intervals indicated by error bars in parentheses. The squared
Mahalanobis distance of this point with respect to the mean and
covariance of the prior is $\mdsquared(\widehat{\lecs}_{\pi
  N},\vec{\mu}_{\pi N};\Sigma_{\pi N}) = 9.65$, which is just far
enough for the posterior MAP value to be outside 95\% of the prior probability
mass. This can be cast in terms of a $p$ value (0.047) and traditional
significance testing, which would lead us to reject the correctness of
the Roy-Steiner prior (null hypothesis) on the 5\% significance
level. However, we are inclined to place doubt on our model for the
truncation error and in particular its variance. As a side remark, the
$\mdsquared$ values for the MAP values of the $\pi N$ LEC posteriors
for $\Delta$-less \nnlo~\cite{Svensson:2021lzs} and for the next order
\nnnlo~\cite{Svensson:2022kkj}, inferred using an uncorrelated
truncation error, are even greater and again point to significant
tension between the respective LECs inferred using \NN\ data and $\pi
N$ data.

We suspect that the truncation error is underestimated since the
\nlo-\nnlo\ shift in the $\Delta$-full theory is on the smaller side,
which is not too surprising as the inclusion of the $\Delta$-isobar
pulls down higher-order contributions in the $\Delta$-less theory. Also, we
cannot rule out that the contributions at $\Delta$-full \nnnlo\ are
substantial owing to the introduction of a rich fourth-order contact
potential. To shed some light on the possible underestimation of
$\cbarsq$ we sampled the joint posterior
$\prob(\lecs,\cbarsq|\data,I)$ in the uncorrelated limit of the
truncation error. For this, we employed the same LEC prior as
defined in Sec.~\ref{sec:prior}, and assumed the same $\cbarsq$ for all
observable types. For the latter parameter we employed an
inverse-$\chi^2$ prior with $\nu=23.75 $ and $\tau^2=0.19$, as
obtained via conjugacy of an inverse$-\chi^2$ (hyper)prior with
$\nu_0=\tau_0^2=1$ updated using training data from expansion
coefficients $c^{(3)}$ for
$\sigma(\theta),A_y(\theta),\sigma_\text{tot}$ on the same grid of
$\theta$ and $\tlab$ values as used when inferring the $\GP$ model
parameters for the correlated truncation error. This prior is sharply
peaked at $\cbarsq\approx 0.45^2$. The marginal LEC values from the
$\prob(\lecs,\cbarsq|\data,I)$ posterior are not noticeably different
from before. However, the posterior modes for $\cbarsq$ are $0.96(5)$
and $3.5(2)$ at \nlo, and \nnlo, respectively, with 68\% credible
intervals indicated in parentheses. Clearly, conditioning on $\data$
has a significant impact and increases the EFT truncation error.

\subsection{Posterior predictive distributions}

In this section we quantify the posterior predictive distributions
(\ppd s) for selected \np\ scattering observables at \nlo\ ($k=2)$ and
\nnlo\ ($k=3$).
A PPD is a distribution of values for a, possibly unseen, observable
$\ystar$ conditioned on experimental data~\cite{bda3}.
Specifically, we will sample from $\prob(\ystar | \data, \idptstar,
I)$ with $\ystar = \yth(\idptstar) + \dyth(\idptstar)$ at some
value(s) $\idptstar$ for the independent variable.
Note that we include our estimate of the truncation error in all \ppd s. To achieve
this, we sample $\yth^{(k)}(\lecs;\idptstar)$ for $\lecs \sim
\prob(\lecs|\data,I)$, i.e., we draw LEC values from the MCMC chains
distributed according to the posteriors quantified in
Sec.~\ref{sec:sampling}. To each sample we add a normally distributed
truncation error $\dyth^{(k)}(\idptstar) \sim
\mathcal{N}(0,\Sigma^{(k)}_{\text{th},y})$ where the covariance matrix
is informed by a $\GP$ with MAP values
$\widehat{\lsth},\widehat{\lsE},\widehat{\cbarsq}$ according to the
values listed in the table in Fig.~\ref{tab:database}.

\begin{figure}[tb]
\centering
\includegraphics[width=1.0\linewidth]{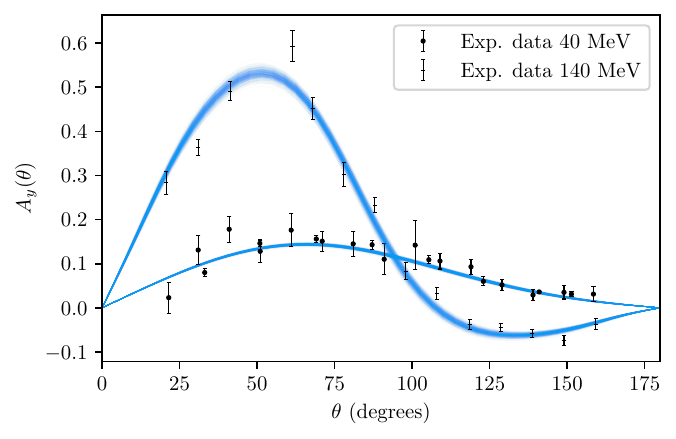}
\caption{500 draws from the \nlo\ \ppd s for $A_y(\theta)$ at
  $\tlab=40$ MeV and $\tlab=140$ MeV using a correlated truncation
  error. The experimental data at $40$ and $140$ MeV are
  from Refs.~\cite{Wilczynski:1984,Langsford:1965} and
  \cite{Stafford:1957}, respectively.}
\label{fig:ppds_P_40_140MeV}
\end{figure}

As a model check we first inspect the \ppd\ for some of the seen data,
i.e., the data already in $\data$. In Fig.~\ref{fig:ppds_P_40_140MeV}
we show 500 draws from the \ppd s for the vector polarization
$A_y(\theta)$ at \nlo\ with $\tlab=40$ MeV and $\tlab=140$ MeV.
\begin{figure*}[t]
\centering
\includegraphics[width=1.0\linewidth]{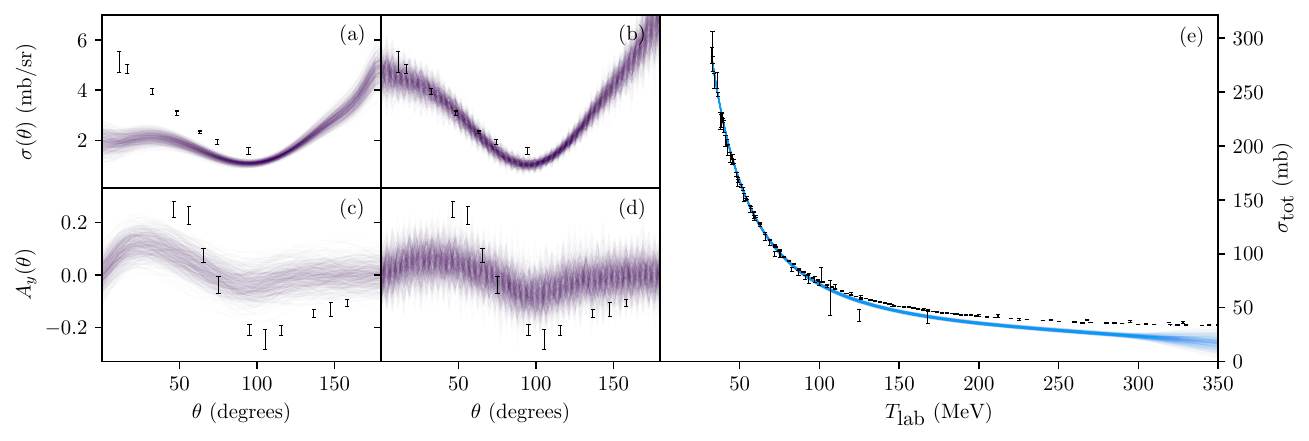}
\caption{500 draws from the \ppd s for unseen data for: (a)
  $\sigma(\theta)$ at \nnlo\ and $\tlab=319$ MeV with a correlated
  error, and experimental data from Ref.~\cite{Keeler:1982}. (b) Same
  as (a) but with an uncorrelated error.  (c) $A_y(\theta)$ at
  \nnlo\ and $\tlab=350$ MeV with a correlated error, and experimental
  data from Ref.~\cite{Siegel:1956}. (d) Same as (c) but with a
  correlated error. (e) $\sigma_\text{tot}$ at \nlo\ and
  $\tlab=30$-$350$ MeV (see Ref.~\cite{perez13-2} for references to the
  experimental data).}
\label{fig:ppds_SGT_DSG_P}
\end{figure*}
The model calculations were performed using EC emulators trained as
above. The observed correlation length, $\widehat{\lsth}=28$ degrees,
is reasonable, and we hit most of the data within the theory
uncertainty. Due to the symmetries of the strong interaction we must
have $A_y(\theta=0)=A_y(\theta=180)=0$. As we incorporated this type
of constraint during the training the $\GP$, the predictive variance
indeed goes to zero for $\theta=0,180$ degrees. Given the rather long
correlation lengths, the variance constraints at the angular endpoints
appears to propagate to the interior to further suppress the
truncation error. We have not studied the empirical coverage of the
truncation error as we expect it to extrapolate rather poorly to
unseen data at higher energies.

In Fig.~\ref{fig:ppds_SGT_DSG_P} we quantify \ppd s for unseen data
and compare predictions based on correlated and uncorrelated
truncation errors. We have drawn 500 samples from the \ppd s for the
differential cross section $\sigma(\theta)$ at $\tlab=319$ MeV, the
vector polarization $A_y(\theta)$ at $\tlab=350$ MeV, and the total
cross section $\sigma_\text{tot}$ for $\tlab=30-350$ MeV, where the
latter is at \nlo\ and the former two at \nnlo\ using correlated as
well as uncorrelated truncation errors.

The smoothing effect of the correlated truncation error is clearly
visible in panels (a) and (c). For $\sigma(\theta)$, in panels (a) and
(b), the predictions in the low-$\theta$ region fare significantly
worse when including correlations. We speculate that this is connected to the reduced weight of high-energy \np\ data residing within one correlation length of the truncation error. In panels (c) and (d) we compare
$A_y(\theta)$ predictions at \nnlo. Besides overall similarities, the
predictive variance of $A_y(\theta$) is finite at the edges even when
we include a correlated truncation error for which we have imposed the
relevant symmetry constraints. However, at $\tlab=350$ MeV we are
approximately one correlation length away from the training point
where the constraint was imposed, and its effect has deteriorated.
Finally, the \ppd\ for the total cross section in panel (e)
underestimates the experimental data for $\tlab>100$ MeV, and the
correlated truncation error appears to be somewhat too small as well.

\section{Summary and Outlook}
\label{sec:summary}
We quantified a correlated truncation error for $\Delta$-full
\chieft\ predictions of \np\ scattering observables at \nlo\ and
\nnlo. The correlation structure was modeled using a $\GP$ with a
squared-exponential kernel, and the resulting MAP values for the
correlation lengths in the scattering energy and angle directions are
in the ranges of 45--83 MeV and 24--39 degrees, respectively, for the
17 different observable types in the set of \np\ scattering data that
we consider. These are significant correlation lengths that, in
principle, could reduce the effective number of independent \np\ data in the
 likelihood by two orders of magnitude and therefore strongly impact the LEC inference. However, other than doubled widths of the univariate marginal posteriors for all
LECs we find that the introduction of a correlated EFT
truncation error does not change the structure of the LEC posteriors by much. This is  explained well by the relatively
small marginal variances of the truncation errors that we quantify in this paper. Indeed, we found effective
dimensions of the data likelihood that were approximately 1/8 and 1/4 of the
length of the experimental \np\ database, at \nlo\ and \nnlo\ respectively.

The marginal variance of the uncorrelated truncation error increases
up to four times when jointly sampling it with the LEC values. Thus,
future inferences should attempt to marginalize predictions over the
hyperparameters of the $\GP$ model and if possible also the
breakdown scale, or the expansion parameter $Q$. For this, the HMC
algorithm with AD should provide the necessary leverage to enable
sampling of a
posterior with, at least, doubled dimensionality. There is most
likely useful information to learn about the truncation error from
theoretical predictions at \nnnlo\ in $\Delta$-full
\chieft. Adding more chiral orders in the analysis of the truncation error might reveal challenging structures in the order-by-order data that call for more sophisticated $\GP$-kernel design than we employ in this paper.

It will be interesting in future studies to explore the performance of
the inferred interactions in predictions of many-nucleon observables.
The general \emph{ab initio} modeling capabilities of
$\Delta$-full chiral interactions at \nnlo{} were studied recently in
Refs.~\cite{Hu:2021trw,Jiang:2022tzf,Jiang:2022oba}. These papers
targeted, in particular, predictions for heavy nuclei up to $^{208}$Pb
and infinite nuclear matter. A common strategy was to first use
history matching to identify
interaction parametrizations that give acceptable model predictions
for a suite of low-energy nuclear observables. Energies and radii of
light nuclei, as well as selected \np{} phase shifts, were included
sequentially in an implausibility measure that led to the removal of
large regions of the LEC parameter space.
Although a lower regulator cutoff $\Lambda=394$~MeV was used in the
history-matching analyses, it is still interesting to compare the
identified regions with the parameter posterior \pdf{} inferred in
this paper. First, we find
that the posterior mode from this paper projected on $c_{2,3,4}$ is
situated just outside the non-implausible range. That is not
surprising since the Roy-Steiner prior was employed without updates throughout history
matching. Furthermore, the mode is located at the upper end of the
non-implausible range in the $C_{1S0}$ direction and at the lower
edges of the $C_{1P1}$ and $C_{3P2}$ ranges. The significance and
origin of this tension should be explored in a joint analysis of \NN,
few-, and many-nucleon observables.

The scattering emulators and HMC sampling presented here can be
combined with history matching and reduced-basis methods for many-body
observables to enable a joint analysis of chiral interactions. Such
studies, however, will rely on an improved understanding of correlated
EFT truncation errors and soft momentum scales in finite nuclei.

\begin{acknowledgments}
We thank Jordan Melendez for providing detailed and insightful comments
on the manuscript.
This work was supported by the European Research Council (ERC) under
the European Unions Horizon 2020 research and innovation program
(Grant Agreement No. 758027), the Swedish Research Council (Grant
No. 2021-04507). The computations were enabled by
resources provided by the Swedish National Infrastructure for
Computing (SNIC) partially funded by the Swedish Research Council
through Grant Agreement No. 2018-05973.
\end{acknowledgments}

\bibliography{bibliography.bib}

\appendix

\section{LEC posteriors \pdf s}
\label{app:posteriors}
\begin{figure*}[t]
\centering
\includegraphics[width=1.0\linewidth]{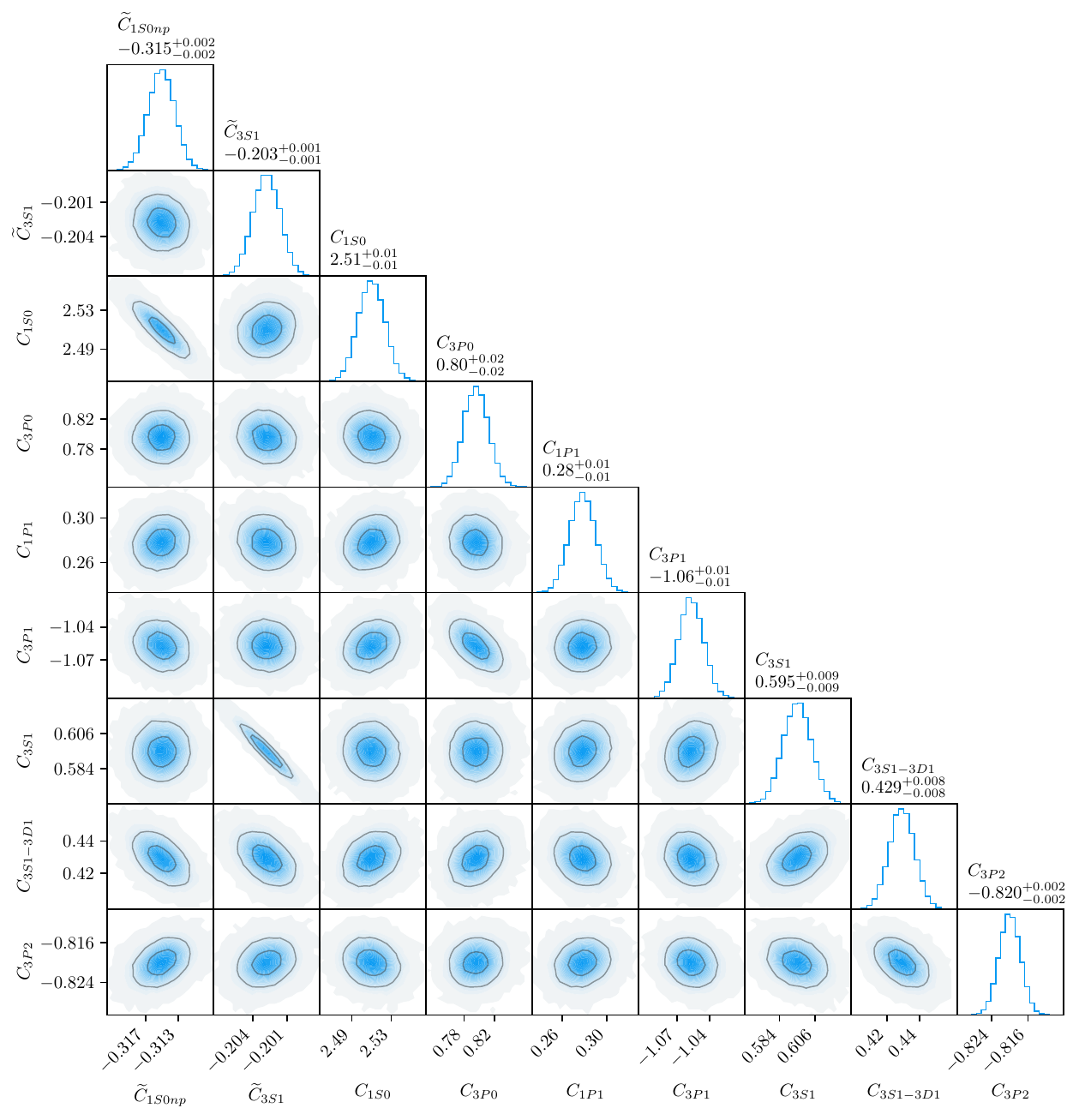}
\caption{Posterior \pdf\ of the values for the LECs at \nlo\ inferred
  using an EFT truncation error in the uncorrelated limit. The inner
  (outer) gray contour line encloses 39\% (86\%) of the probability
  mass. The dot-dashed vertical lines indicate a 68\% credibility
  interval in the univariate marginals. See main text for units and
  conventions.}
\label{fig:nlo_uncorrelated}
\end{figure*}

\begin{figure*}[t]
\centering
\includegraphics[width=1.0\linewidth]{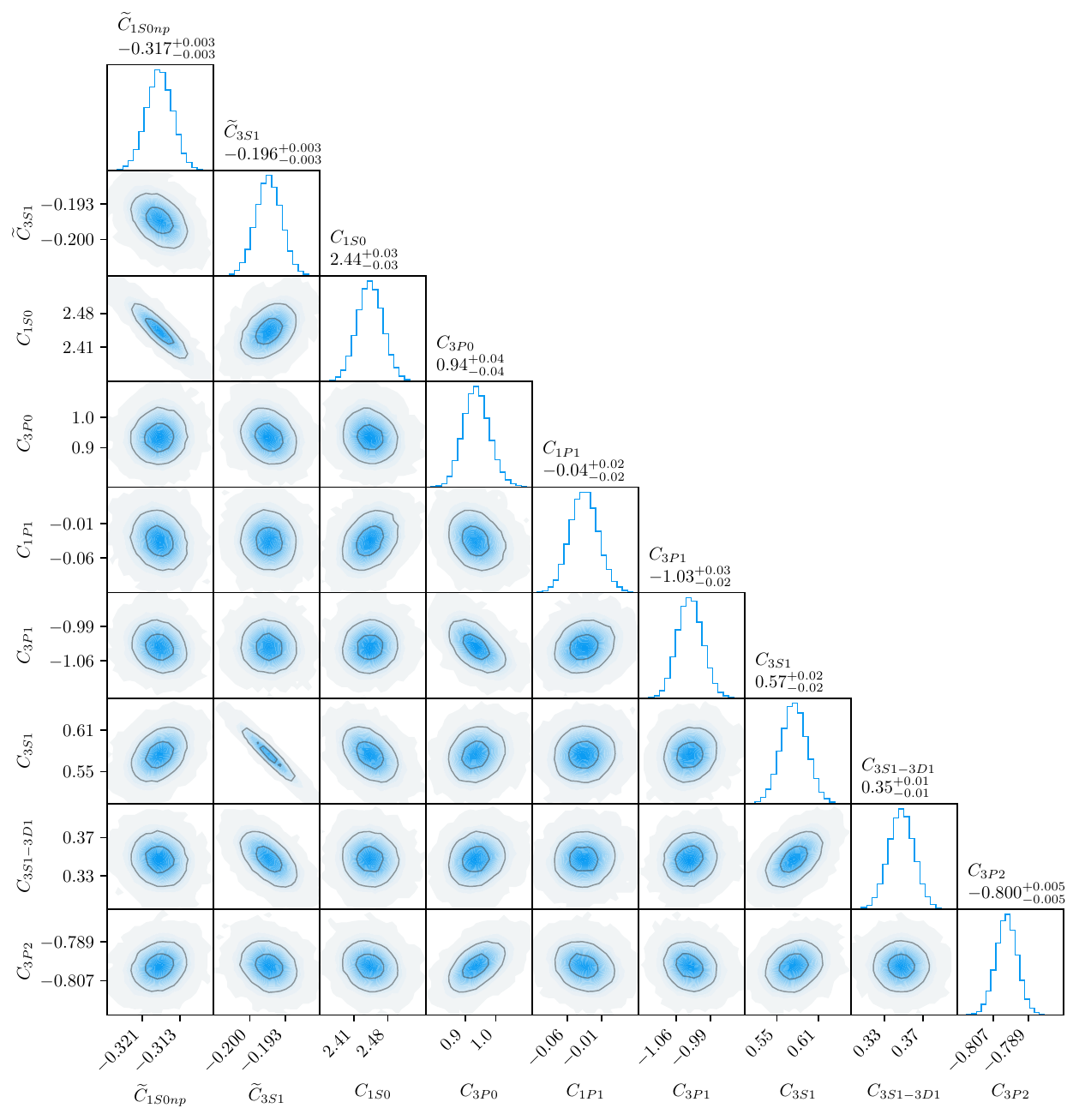}
\caption{Posterior \pdf\ of the values for the LECs at \nlo\ inferred
  using a correlated model for the EFT truncation error. See
  Fig.~\ref{fig:nlo_uncorrelated} for figure notation.}
\label{fig:nlo_correlated}
\end{figure*}

\begin{figure*}[t]
\centering
\includegraphics[width=1.0\linewidth]{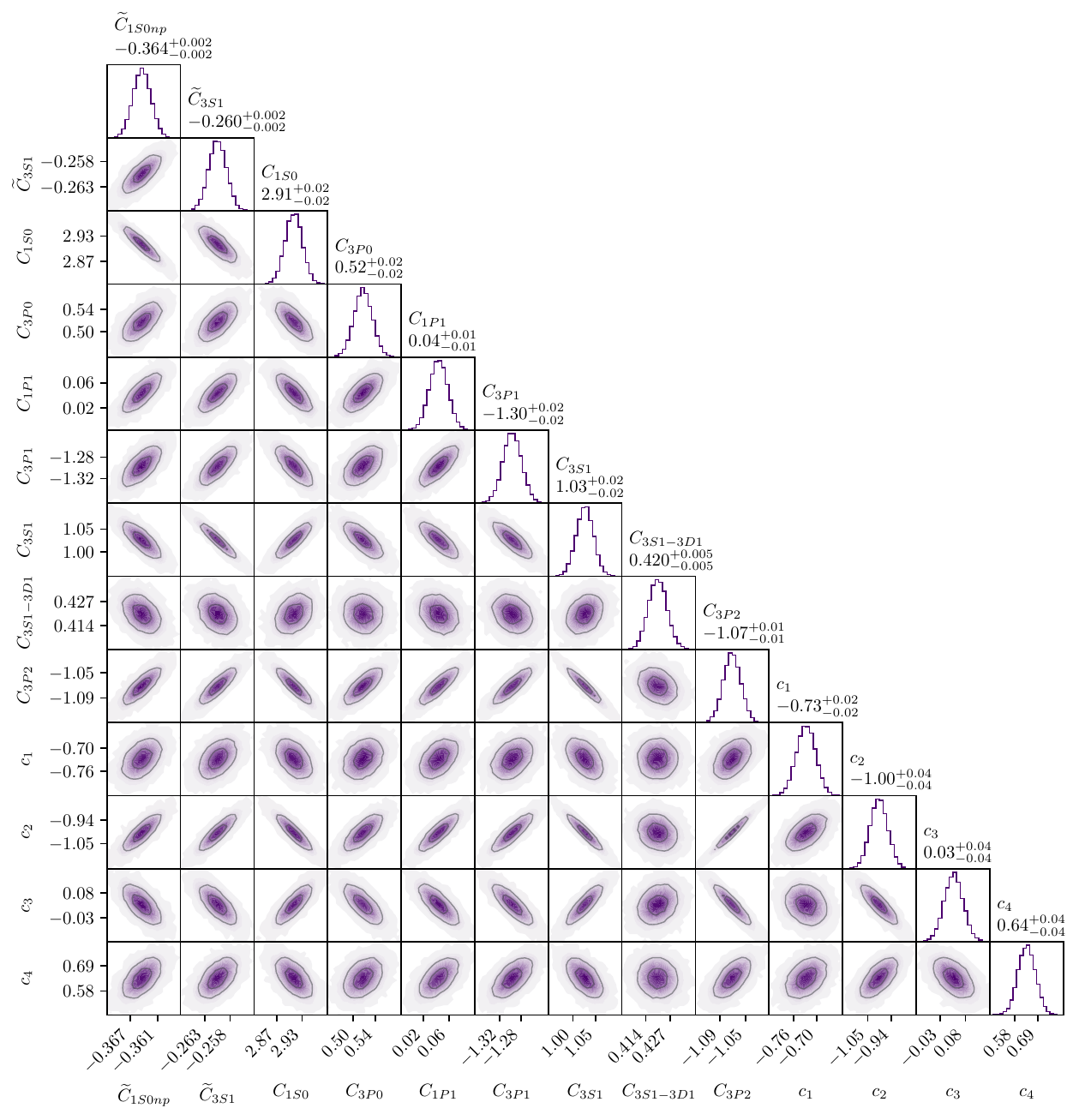}
\caption{Posterior \pdf\ of the values for the LECs at \nnlo\ inferred
  using an EFT truncation error in the uncorrelated limit. See
  Fig.~\ref{fig:nlo_uncorrelated} for figure notation.}
\label{fig:nnlo_uncorrelated}
\end{figure*}

\begin{figure*}[t]
\centering
\includegraphics[width=1.0\linewidth]{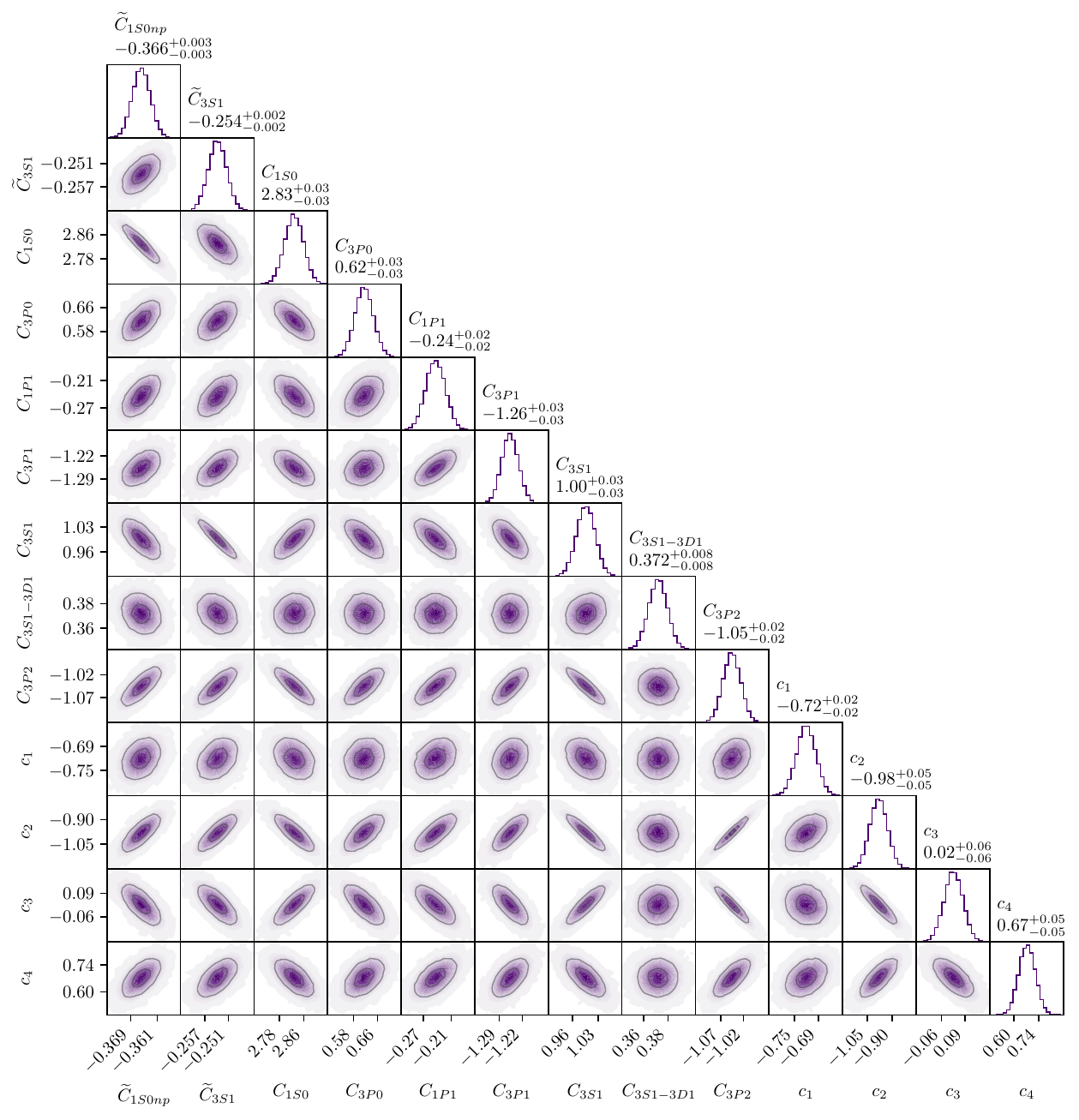}
\caption{Posterior \pdf\ of the values for the LECs at \nnlo\ inferred
  using a correlated model for the EFT truncation error. See
  Fig.~\ref{fig:nlo_uncorrelated} for figure notation.}
\label{fig:nnlo_correlated}
\end{figure*}

In this appendix we show (Figs.~\ref{fig:nlo_uncorrelated}--\ref{fig:nnlo_correlated}) the LEC posterior \pdf s for the values of
the LECs in $\Delta$-full \chieft\ up to \nnlo, conditioned on
\np\ scattering data with $30 \leq \tlab \leq 290$ MeV and a
truncation error in the correlated and uncorrelated limits. The corresponding MCMC chains are available upon request.

\end{document}